\documentclass[APA,LATO1COL]{WileyNJD-v2}

\usepackage{mathtools}
\usepackage{float}

\algnewcommand\algorithmicinit{\textbf{Initialize:}}
\algnewcommand\Init{\item[\algorithmicinit]}

\articletype{Article Type}%

\received{26 April 2022}
\revised{6 June 2016}
\accepted{6 June 2016}

\begin{document}


\title{Hierarchical Embedded Bayesian Additive Regression Trees}

\author[1,2]{Bruna Wundervald*}

\author[1,2]{Andrew Parnell}

\author[1,2]{Katarina Domijan}

\authormark{Wundervald \textsc{et al.}}

\address[1]{\orgdiv{Hamilton Institute}, \orgname{Maynooth University}, \orgaddress{\state{Kildare}, \country{Ireland}}}

\address[2]{\orgdiv{Maths \& Stats Department}, \orgname{Maynooth University}, \orgaddress{\state{Kildare}, \country{Ireland}}}

\corres{*Bruna Wundervald, \email{brunadaviesw@gmail.com}}

\presentaddress{Maynooth, Kildare, Ireland.}

\abstract[Summary]{We propose a simple yet powerful extension of Bayesian Additive Regression Trees which we name Hierarchical Embedded BART (HEBART). The model allows for random effects to be included at the terminal node level of a set of regression trees, making HEBART a non-parametric alternative to mixed effects models which avoids the need for the user to specify the structure of the random effects in the model, whilst maintaining the prediction and uncertainty calibration properties of standard BART. Using simulated and real-world examples, we demonstrate that this new extension yields superior predictions for some of the standard mixed effects models' example data sets, and yet still provides consistent estimates of the random effect variances. We illustrate the new tool with example data sets from widely-used mixed effects model packages. 
}


\jnlcitation{\cname{%
\author{Wundervald B.}, 
\author{A. Parnell}, and 
\author{K. Domijan}} (\cyear{2022}), 
\ctitle{Hierarchical Embedded Bayesian Additive Regression Trees}, \cjournal{Stat}, \cvol{2022;000}.}

\maketitle

\footnotetext{\textbf{Abbreviations:} BART, Bayesian Additive Regression Trees; HEBART, Hierarchical Embedded Bayesian Additive Regression Trees; MH, Metropolis-Hastings; MCMC, Monte Carlo Markov Chain; RMSE, Root Mean Squared Error; LME, Linear Mixed-Effects;}

\section{Introduction}\label{sec1}

Bayesian Additive Regression Trees \citep[BART;][]{chipman2010bart} is a commonly-used probabilistic machine learning approach that produces predictions based on sums of regression trees.
Like most standard machine learning approaches, however, BART is not
mathematically designed to deal with hierarchical structures in the data. For example, certain observations may share common grouping characteristics e.g. repeated measures and grouped data \citep{gelman2006data},
where there is an intra-group
variance that needs to be accounted
for when fitting the model. 
There are, of course, some algorithm
options that can fit statistical methods
to grouped data \citep[e.g.][]{bates2011package}, 
but they are often 
not flexible or able to adapt to 
data that has been generated by
a complicated underlying structure. 
It
might also happen that certain 
observations have their grouping
information missing 
\citep{vallejo2011comparison}, 
or predictions might be required at different levels of the data hierarchy, which 
leads to even more complicated scenarios. 
Occasionally it is of interest to estimate and/or remove the variability associated with the structured component of the data. The user is often left with a difficult model selection task, as they must choose whether the random effects go into an intercept or a slope or an interaction. Due to the complexity of this task, the model space is not usually well-explored.  Literature has arisen on the need to explore and check hierarchical models  \citep[see e.g. Chapter 24 of][]{gelman2006data}. 

With this in mind, Hierarchical Embedded BART (HEBART) introduces a hierarchical component in order
to properly
model grouped data and
 estimate the manner through which the hierarchical component of the data enters into the predictions using Bayesian Additive
Regression Trees. 
This is substantially different to the original means by which random effects have been proposed as an inclusion to the BART model; see Section 7 of \cite{chipman2010bart}. By contrast, in the simplest version of our model, the hierarchical component is a single categorical predictor variable that adjusts all the predictions in the terminal nodes of each tree. Thus the effects are no longer confined to individual components of the model (e.g. intercept terms) and so, give much greater flexibility whilst still providing the predictions at multiple layers of the hierarchy. Moreover, we do not
 make it a requirement for the grouping information
 to exist when calculating the predictions, 
 which allows us to predict for observations
 where the group is missing, or when the variability associated with the grouping information should be removed, bringing
 even more flexibility to our approach. 
 We provide full technical details of our extension below. 

Our paper is structured as follows. In Section 2 we introduce the BART model mathematically, and discuss the fitting algorithm and some extensions that have already been proposed. In Section 3 we outline our new HEBART approach and discuss how this extends BART into a generalized model for hierarchical data structures, whilst retaining the attractive properties and algorithmic efficiency that BART exhibits. In Section 4 we demonstrate the performance of the method on simulated and real-world data. Finally, Section 5 discusses some of the potential future research areas and drawbacks of our approach. An appendix contains some of the more detailed mathematics behind the fitting algorithm. 

\section{Introduction to Bayesian Additive Regression Trees}\label{sec2}

\subsection{The BART model}

The initial Bayesian Regression Tree model was first
proposed over 20 years ago \citep{chipman1998bayesian}, 
and consists of an algorithm that fits CART decision trees using Bayesian inference. The same authors extended this
method to create the Bayesian Additive Regression Tree
\citep{chipman2010bart} approach, which assumes that the generating
system of a continuous random variable 
$\mathbf{y} = [ y_1, \dots, y_n]$ can be approximated by a 
sum of regression trees. In a standard regression setting the BART model is usually written as:

\begin{equation}
    y_i = \sum_{p = 1}^{P} \mathbf{G}(X_{i}; \mathcal{T}_{p}, \Theta_{p}) + \epsilon_{i}, \;  \epsilon_{i} \sim N(0, \tau^{-1}), 
\label{eqn_bart}
\end{equation}

\noindent for observations $i = i, \dots, n$, and $p = 1,\ldots,P$ trees with $P$ the total (fixed) number of trees, $X_{i}$ represents the set of covariates; $\mathcal{T}_{p}$ is a tree structure, and $\Theta_p$ represents a set of terminal node parameters. The function $\mathbf{G}$ returns a terminal node prediction from $\Theta$ by passing $X_i$ through the tree $T_p$. $\Theta_{p}$ consists of a set of $\mu_{p,l}$ parameters for each of the $l= 1, \dots, L_p$ 
terminal nodes in tree $p$. These values provide the tree-level predictions which are summed together to give an overall predicted value. The residual term $\epsilon$ is assumed normal with residual precision $\tau$. Figure \ref{diagram} (left panel) shows a standard single tree that a BART model may use. 

We write the set of all trees and parameters as $\mathcal{T}$ and $\Theta$ respectively. The joint posterior distribution of the trees and all the parameters is then given by:

\begin{equation}
P(\mathcal{T}, \Theta, \tau | \mathbf{X}, \mathbf{y})   \propto
\bigg[ \prod_{p = 1}^{P} \prod_{b = 1}^{L_{p}} \prod_{i: \mathbf{x}_i \in \mathcal{D}_{p, b}}
p(y_i | \mathbf{x}_i, \mathcal{T}_{p}, \Theta_p, \tau) \bigg] 
\times \bigg[ \prod_{p = 1}^{P} \prod_{b = 1}^{L_{p}}  p(
\mu_{p, b} |  \mathcal{T}_{p}) p( \mathcal{T}_{p}) \bigg] p(\tau),
\label{eqn_posterior}
\end{equation}

\noindent where $p(y_i | \mathbf{x}_i, \mathcal{T}_{p}, \Theta_p, \tau)$ is the normally distributed likelihood as defined in Equation \ref{eqn_bart}, $\mathcal{D}$ represents
the regions in the covariates space (e.g. tree nodes), and $b$ indexes the terminal nodes in tree $p$. The term $p(\mu_{p, b} |  \mathcal{T}_{p})$ is the prior distribution on the terminal node parameters across each terminal node $b$ in each tree $p$. $p( \mathcal{T}_{p})$ is the prior distribution on the tree structure, and $p(\tau)$ is the prior on the residual precision. We detail all these terms below.

The prior distribution on the trees proposed by \citet{chipman2010bart} involves applying a separate term for each node in the tree and considering both the probability of a split as well as the probability of a new splitting variable being chosen. For an entire tree $\mathcal{T}_p$, we have: 

$$P(\mathcal{T}_p) \propto \prod_{\eta \in L_{\mathcal{T}}} (1 - P_{SPLIT}(\eta ))\prod_{\eta \in L_{\mathcal{I}}}
P_{SPLIT}(\eta ) 
\prod_{\eta \in L_{\mathcal{I}}} 
P_{RULE}(\rho | \eta ). $$

\noindent where $L_{\mathcal{T}}$ and $L_{\mathcal{I}}$ represent the
sets of terminal and internal nodes, respectively, $\rho$ represents a generic splitting value, and $\eta$ represents a node in the tree. The probability of a node being non-terminal is given by $P_{SPLIT}(\eta) = \alpha(1 - d_{\eta})^{-\beta}$,  
where $d_{\eta}$ denotes the depth of the node $\eta$. The recommended values for the hyperparameters are $\alpha \in (0, 1)$ and $\beta > 0$, which control the depth and bushiness of the trees. For the  probability of the new splits, 
$P_{RULE}(\rho | \eta, \mathcal{T}) = \frac{1}{p_{adj}(\eta)} \frac{1}{n_{j.adj}(\eta)}$
 where $p_{adj}(\eta)$ represents how many predictors are still
available to split on in node $\eta$, and $n_{j.adj}(\eta)$ how many values in a given predictor are still available. 

The prior distribution for the terminal node and overall parameters in the standard regression case is denoted by  $p(\Theta, \mathcal{T})$. This is given a prior with a standard
conjugate form:

$$\mu_1, \dots, \mu_{L} | \tau_{\mu}, \mu_\mu, \mathcal{T} 
\sim \mathcal{N}(\mu_\mu, \tau_{\mu}^{-1}),$$
  
\noindent where $\tau_{\mu}^{-1}$ and $\mu_{\mu}$ are chosen
such that a high density of this distribution
is apportioned to the range $[y_{min}, y_{max}]$ interval,
by setting $P \mu_{\mu} - k \sqrt{P(\tau_{\mu}^{-1})} = y_{min}$ and $P \mu_{\mu} + k \sqrt{P(\tau_{\mu}^{-1})} = y_{max}$, for some value of $k$. The response $y$ is usually standardized before the model is run which allows for reasonable guesses as to the hyper-parameter values of $\mu_\mu$ and $\tau_\mu$, though these too can be estimable parameters. 

The residual precision prior is set as 
$\tau \sim  \text{Gamma}(\nu/2, \gamma\nu/2),$ where $\gamma$ and  $\nu$ are fixed. An oft-used tactic is to set the two hyper-parameters such that the BART residual precision is greater than an equivalent precision value from a standard linear regression model applied to the same data with a high probability. We follow the same guidance in our extension to the model as outlined below. 

Since its creation, BART has been applied in a wide variety of different application areas. The model has been shown to be useful for
credit risk modeling
\citep{zhang2010bayesian}, survival data analysis \citep{sparapani2016nonparametric, sparapani2020nonparametric}, ecology and evolution
modelling \citep{carlson2020embarcadero}, weather and avalanche forecasting
\citep{blattenberger2014avalanche}, 
and genetics \citep{waldmann2016genome}. A popular approach
is its use in causal inference \citep{hill2011bayesian, hahn2020bayesian}, 
where BART produces accurate estimates of average treatment effects and is competitive even with the true 
data generating model. 

Beyond applications, many fundamental extensions to the standard BART model have been proposed. Some of the first include adapting
BART for categorical, count, and multinomial regression 
\citep{murray2017log, kindo2016multinomial} and quantile regression \citep{kindo2016bayesian}. This was followed
by the proposal of BART that adapts to
smoothness and sparsity \citep{linero2018bbayesian}, 
models for high-dimensional data and variable
selection \citep{linero2018abayesian}, 
BART for zero-inflated and semi-continuous responses
\citep{linero2020semiparametric} and
an extension proposed by 
\citep{hernandez2018bayesian}, where the
authors combine BART with Bayesian Model Averaging to obtain posterior distribution more efficiently when there is a large number of 
variables available. 
More recently BART has been extended for use with
heterocedastic data \citep{pratola2020heteroscedastic}, 
for the estimation of monotone and smooth surfaces, 
 \citep{starling2020bart},  
varying coefficient models 
\citep{deshpande2020vcbart}, semiparametric BART
\citep{prado2021semi}, and a combination of BART with model-trees \citep{prado2021bayesian}. As this is 
a fairly new class of machine learning algorithm, progress on the theoretical performance of BART has only just begun. Some of the mathematical properties of 
BART, including a deep review of the BART methodology can be found in 
\citet{linero2017review}, and some more general
theoretical results in
\citet{rovckova2020posterior, rovckova2019theory}. 

\subsection{Fitting algorithm}

The BART model is fitted via a  backfitting MCMC algorithm
\citep{hastie2000bayesian} that 
holds all other trees constant while the current one is being updated. 
This involves calculating the full conditional distribution
$P(\mathcal{T}_{p}, \Theta_{p} |  \tau, \mathbf{X}, \mathbf{y}, \mathcal{T}_{(p)}, \Theta_{(p)})$, where $\mathcal{T}_{(p)}$ represents the set of all trees \textit{except}
for tree $p$ (the definition of $\Theta_{(p)}$ is analogous). This conditional depends on $(\mathbf{X}, \mathbf{y}, \mathcal{T}_{(p)}, \Theta_{(p)})$ only via the current state of the residuals, defined as 

$$ R_{i,p} = y_i - \sum_{l \neq p} \mathbf{G}(X_i; \mathcal{T}_{l}, \Theta_{l}),$$

\noindent meaning these partial residuals include the sum of the predictions for all trees except $p$. The choice of prior distributions on the trees and terminal nodes allows for the term $P(R_p |\mathcal{T}_{p}, \tau)$ to be calculated in closed form which avoids the need to trans-dimensional MCMC methods, and greatly simplifies the resulting algorithm.  

The algorithm samples from $P(\mathcal{T}_{p}, \Theta_{p} | \tau, R_p) $
via two main steps:

\begin{enumerate}
    \item propose a new tree through one of 4 proposal moves and calculate
    $P(\mathcal{T}_{p} | \tau, R_p) \propto  P(\mathcal{T}_{p}) P(R_p |\mathcal{T}_{p}, \tau)$, and
    \item sample a new set of terminal node parameters via $P(\Theta_{p} |R_p, \mathcal{T}_{p}, \tau)$, for the new tree
\end{enumerate}

The new trees are proposed using a Metropolis-Hastings sampler \citep{brooks2011handbook}, where the candidate tree is obtained via: GROW (a terminal node is selected uniformly from the set of terminal nodes and split into two, with a new split variable and split value chosen analogously), PRUNE (a pair of terminal nodes with a common parent are collapsed together), CHANGE (a splitting rule is chosen uniformly across the tree and is changed to a new split variable and split value) or SWAP (a parent-child pair of internal nodes is chosen uniformly and swapped in the tree). The movement choice requires probabilities for each move, which can be equal or depend on prior beliefs about moves that should be prioritized. The means by which the splitting variable and value are chosen are slightly different across the implementations of BART in the literature. \citet{pratola2016efficient} propose even more detailed moves for the generation of new candidate trees, and \citep{linero2018bayesian} puts a Dirichlet hyperprior on the splitting rules, such
that the algorithm keeps track of which variables
are most useful for splits and avoid using too 
many variables. 

\section{Hierarchical Embedded Bayesian Additive Regression Trees (HEBART)}

\begin{figure}[ht] 
\centerline{\includegraphics[width=310pt, height=180pt]{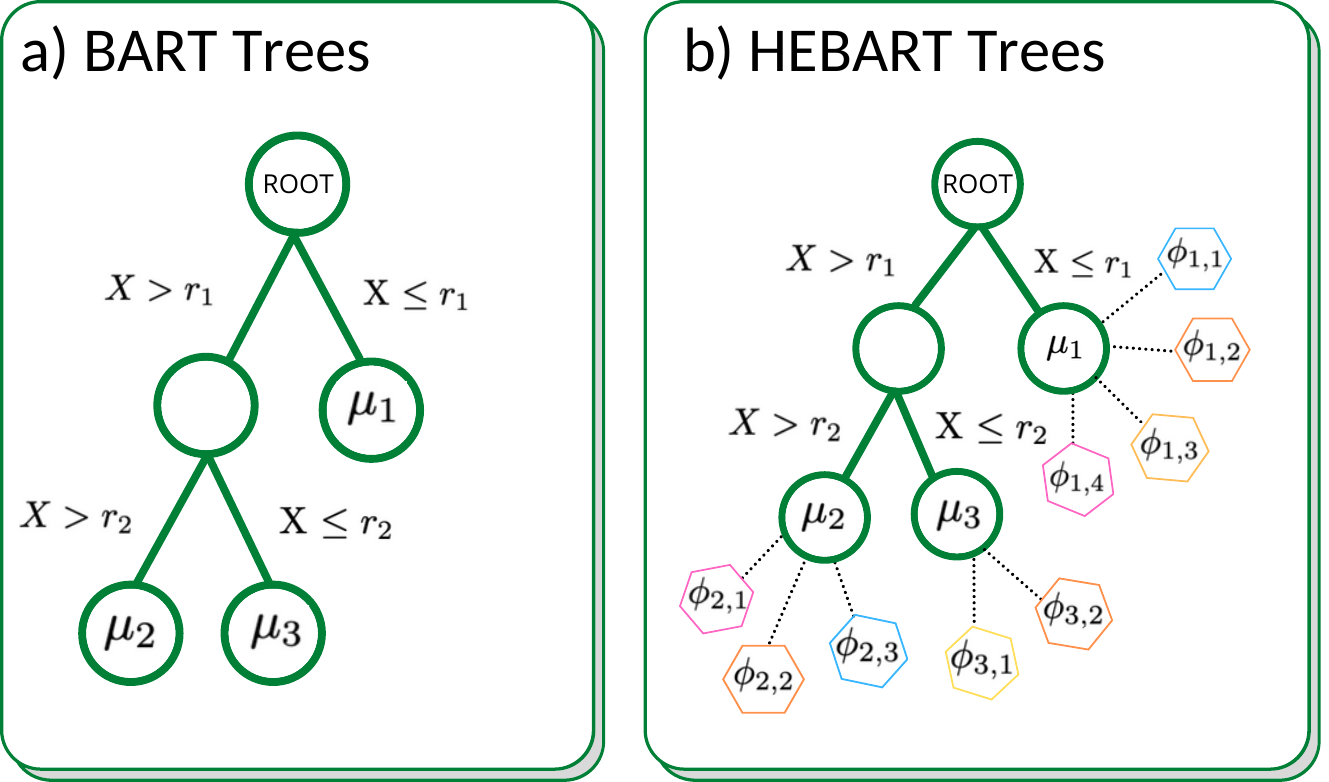}}
\caption{Left panel: a standard BART tree using one covariate $X$. Right panel: a HEBART tree with one covariate $X$ and a single grouping variable with four levels. In BART, 
each terminal node $b$ has only one terminal node
parameter $\mu_b$, represented by 
$\mu_1$, $\mu_2$, and $\mu_3$. In HEBART, each terminal
node $b$ has its own overall $\mu_{b}$ parameter, plus
one $\phi_{b, j}$ for each of the $J$ groups in the node (not all groups need to have associated data in each terminal node). For example, terminal node 2 has an overall parameter
$\mu_2$, and intra-group parameters ($\phi_{2, 1}$, 
$\phi_{2, 2}$, $\phi_{2, 3}$), one for each of the possible groups.
\label{diagram} }
\end{figure}

Our HEBART approach merges the ideas from traditional Bayesian hierarchical modeling \citep{gelman2006data} and linear mixed effects models \citep{pinheiro2000linear} with BART. We allow each tree to have an extra split on each terminal node corresponding, in the simplest version, to a random intercept for each member of a categorical predictor variable $z_i$ which takes values $j = 1,\ldots,J$ according to the group membership of observation $i$. Thus we introduce intra-group node parameters which we write as $\mathbf{\phi}_{b,j}$ as the estimate for group $j$ in terminal node $b$. We refer to these parameters as the sub-terminal node level. 
Here, the term `embedded' is used to represent the inclusion
of the grouping variable into the BART model at the terminal node level   
rather than as a simple addition on the BART mean, as was originally proposed as an extension to BART in \citet{chipman2010bart}.  
The new parameters allow us to have a group-specific prediction for each node, as well as an overall terminal node prediction $\mu$. The flexibility of this structure means that there is no requirement for the user to specify where the random effect is included, for example as an intercept or as a regression slope. With HEBART  we can fit Bayesian Additive Regression Trees to any kind of grouped data where there is such a categorical predictor, such as longitudinal data, repeated measures data, multilevel data, and block designs. In addition, having the two levels of predictions is advantageous for scenarios where the group information is not available for all or a subset of the new data. The Bayesian paradigm allows for imputation of any missing groups at any of the terminal nodes. In Figure \ref{diagram} we show a standard BART tree alongside that of our new HEBART trees.

To define the HEBART model, let $P$ be the number of trees, $J$ to be the total number of groups, and $\Theta_p$ the set of node parameters at both the terminal node and sub-terminal node level. More fully, we have one set $\mu_{p}$ which are terminal node predictions for each tree, and one set $\phi_{p, j}$ for each group within each sub-terminal node for tree $\mathcal{T}_p$, $p = 1, \dots, P$. Assuming we have a continuous variable of interest, the fundamental HEBART appears similar to the standard BART approach:

\begin{center} 
\begin{equation}
y_{i} = \sum_{p = 1}^{P} \mathbf{G}(X_{i}, z_i; \mathcal{T}_{p}, \Theta_{p}) + \epsilon_{i},
\end{equation}
\end{center} 

\noindent for observation $i = i, \dots, n$ with grouping variable $z_i$ taking values from $1$ to $J$. In
this specification, we have that $\mathbf{G}$ is the tree look-up function which allows for predictions based on covariates $X_{i}$ and categorical grouping values $z_i$; by default $\mathbf{G}$ will return a sub-terminal node prediction value $\phi$. If the grouping variable $z_i$ is not provided it will return a terminal node estimate $\mu$, which we may write as $\tilde{G}$ for clarity. 
In other words, $\mathbf{G}$ is the function
that maps each estimated
tree structure to the
corresponding values of the covariates $X_{i}$
and finds in which terminal node 
of the tree each observation falls into (see, e.g, Figure 
\ref{diagram}, right panel, for an example). 
With this, the  $\mathbf{G}$ function uses the 
tree structures to attribute 
the correct predicted value for each observation.  
Similar to an LME model, the parameters $\phi$ associated with the categorical grouping variable $z_i$ provide shifts away from the overall means $\mu$ and are constrained using a normally distributed prior.
As before $\mathcal{T}_{p}$ is the tree structure, 
but now $\Theta_{p}$ contains both the terminal node parameters ($\mu$) and the sub-terminal node group parameters ($\phi$) for tree $p$. As with standard BART, the noise is assumed to be distributed as $\epsilon_{i} \overset{\text{iid}}{\sim} N(0, \tau^{-1})$, where $\tau$ is the overall residual precision. 

The $\mu_{p}$ and $\phi_{p, j}$ sets contain, respectively, the overall 
terminal-node mean parameters and the intra-group sub-terminal node mean parameters. In other words, for each tree we will have one $\mu_{p, L_p}$ set for each of their terminal nodes, and for each group within each terminal node we will have another
set $\phi_{p, L_p, j}, j = 1, \dots, J$.  All parameters
in $\Theta_p$ receive priors and 
have their corresponding posteriors, from which
we sample from in the fitting
algorithm. In Figure \ref{diagram} the terminal circles 
represent the terminal nodes, which all have their
own $\mu$ parameters. In the HEBART tree, however, 
we have the addition of the intra-group parameters, 
represented by the hexagonal symbols at the sub-terminal node level. 

In standard BART a minimum node size is usually set on the number of data points that fall into each terminal node. We retain that restriction in our approach but require no such restriction for the sub-terminal node levels. As shown below, the Gibbs update for the $\phi_j$ terms is still available even when no residuals fall into that particular group so we can still provide group-level predictions which can be summed over trees to produce group-level predictions from every tree.

\subsection{Prior distributions}

Many of the standard BART prior distributions carry over to our situation, which desirably means that certain parametric restrictions in our model will yield an exact BART model. Specifically, we have as usual $\mu \sim N(\mu_\mu, \tau_\mu^{-1})$ for the terminal nodes (ignoring node index subscripts for simplicity). As in standard BART, we standardize $y$ so that $\mu_\mu = 0$ and $\tau_\mu$ is simple to calibrate using the heuristics outlined above. We have $\tau \sim \text{Ga}(\nu/2, \gamma\nu/2)$ for the residual precision. We keep the tree prior and its hyper-parameters $\alpha, \beta$ to control the shape and structure of trees. For the sub-terminal node parameters, we simply set $\phi_j \sim N(\mu, \tau_\phi^{-1}/P)$ which forces each sub-terminal node to vary according to $\tau_\phi$ around the terminal node parameter, scaled by the number of trees. Unlike $\tau_\mu$ we treat $\tau_\phi$ as a parameter to be estimated, and provide more details of this below. 


For HEBART, a single sub-terminal node $b$ in tree $\mathcal{T}_{p}$ has partial residuals:

\begin{equation}
R_{ipb} = y_{i}^{(b)} - \sum_{t \neq p} \mathbf{G}(X_{i}^{(b)}, z_i^{(b)}; \mathcal{T}_{t}, \Theta_t)
\end{equation}

\noindent for observation $i$ with cateogorical variable $z_i$. Now, let $\underset{\sim}{R}_j =  \{R_{ij}, \dots, j = 1,\dots, J \}$ represent the full set of residuals
for those observations where $z_i =j$, where we drop the dependence on terminal node and tree for simplicity of notation, 
and vectorize it again so we can write $R \sim N(M \phi, \tau^{-1} I )$. In this fashion, $R$ represents all observations in that particular terminal node for that particular tree now across all groups, $M$ is a binary matrix that allocates observations in the terminal node to groups, and $\phi$ represents the stacked vector of terminal node parameters for all groups in that terminal node. 
Marginalising 
first over $\phi$ and then over $\mu$,  
we obtain a distribution on the partial residuals as $R \sim MVN(0, \Omega_R)$,
where $\Omega_R = \tau^{-1} I + (P \tau_\phi)^{-1} MM^{T} + \tau_\mu^{-1} 1 1^T$. This variance matrix is symmetric and can be inverted quickly when required using Woodbury and related formulae. 


We give the $\tau_\phi$ parameter, responsible for
capturing the intra-group precision,
a $\text{Gamma}(a, b)$ prior.
Since the value of $\tau_\phi$ is analogous to that of a standard LME model we first fit a random intercept model to the same data using the \texttt{lme4} package
\citep{bates2014fitting} and write this estimated precision as $\hat\tau_{\phi}^{\text{LME}}$. We further extract the variance of this estimate via the
\texttt{parameters} package
\citep{parameters} which we write as $\hat \sigma^2(\hat\tau_{\phi}^{\text{LME}})$. The given mean and variance then provide simple point estimates of $a$ and $b$ using the standard method of moments, 
where we find those values such that
$P(\tau_{\phi} < \hat\tau_{\phi}^{\text{LME}}) = 0.5$.
Our approach is analogous to the argument used in standard BART to set $\tau$ via the residual variance of a linear model fit, though there the BART residual variance is expected to fall below the linear model fit with high probability. To set our prior for $\tau \sim \text{Ga}(\nu/2, \gamma\nu/2)$, we calibrate the hyper-parameter values using the same rule as BART, but via the residual variance of the above \texttt{lme4} model fit rather than the standard linear model.
In this case, our strategy aims to yield a 
high probability that $\tau$ is bigger than 
$\hat\tau^{\text{LME}}$, so we find 
prior hyperparameters such that 
$P(\tau > \hat\tau^{\text{LME}}) = 0.95$.

\subsection{Links with standard hierarchical models}\label{subsec:links}

Conditional on the trees, the model can be written as a standard linear mixed effects model. We abuse the above notation slightly to write:
\begin{equation}\label{eq:marg_y_1}
\mathbf{y} | T,\ldots \sim N(\mathbf{S_1} \boldsymbol{\phi}, \tau^{-1} \mathbf{I})
\end{equation}

\noindent where $\mathbf{y}$ is the vector of all observations and $\mathbf{S_1}$ is a binary matrix that allocates each observation to its correct tree and sub-terminal node. The number of columns of $\mathbf{S_1}$ can be large when the number of trees and/or terminal nodes is large, and changes dimension when the conditioning on the trees is removed. $\boldsymbol{\phi}$ here is the stacked vector of all sub-terminal node parameters sorted over all trees. This parameter too can be marginalized over since, within each tree, all components come from a $N(\mu, \tau_\phi^{-1} / P)$ distribution. This second marginalization yields:
\begin{equation}\label{eq:marg_y_2}
\mathbf{y} | T,\ldots \sim N(\mathbf{S_2} \boldsymbol{\mu}, \boldsymbol{\Psi}_y)
\end{equation}
\noindent where now $\boldsymbol{\mu}$ is the stacked terminal node values across all trees and $\Psi_y = \tau^{-1} \mathbf{I} + (P \tau_\phi)^{-1} \mathbf{S_1 S}_1^T$. $\mathbf{S_2}$ is a binary allocation matrix that allocates each observation to the terminal nodes associated with its trees. With this second marginalization there are several links with standard Bayesian mixed effects models frameworks. The model can be seen as a Gaussian Process with kernel autocorrelation function given by $S_1 S_1^T / P$, or as a standard mixed effects model with $\mathbf{y} = \mathbf{S_2} \boldsymbol{\mu} + \mathbf{S_1} \boldsymbol{\phi} + e$ with a prior on $\phi$ centered on zero. The standard theory of LMEs \citep[e.g.][]{pinheiro2000linear} follows directly, and in the Bayesian framework, the updates for sub-terminal node parameters yields the usual partial pooling estimates exemplified by \citet{gelman2006data}; see updates for $\phi$ below. However, the mixing over the trees makes any further theory considerably more complex, and we leave this
as a challenge for future research.

\subsection{Updating parameters}

We use  
MCMC methods to update the parameters, 
using direct Gibbs sampling for updating the terminal node and residual precision parameters, and Metropolis updates for those for which a tractable Gibbs update is unavailable. The updates for an individual tree $\mathcal{T}$ arise by considering the distribution of the partial residuals in a single terminal node, whilst integrating out the mean and variance parameters. We use the marginalized version of $R$ above to find an MH update for a single terminal node in a tree (on the log scale, and summed over terminal nodes $b$) as:

$$\log(\pi(R_1, \ldots, R_{n})) \propto \sum_{b=1}^{L} \left[ -\frac{1}{2} \log | \Omega_{R,b} | -\frac{1}{2} R_b^T \Omega_{R,b}^{-1} R_b \right]$$

For $\mu$, we use the non-marginalised prior for $R$ and the prior on $\mu$ to obtain the full conditional:

$$\mu | \ldots \sim N\left( \left(1^T \Psi_R^{-1} 1 + \tau_\mu\right)^{-1} 1^T \Psi^{-1}_{R} R, \left(1^T \Psi_R^{-1} 1 + \tau_\mu\right)^{-1} \right), $$ which is used to directly sampled values
for the $\mu$ parameters. In this equation, 
we have that $\Psi_R = \tau^{-1} I + (T \tau_\phi)^{-1} MM^{T}$. Similarly, we
also obtain a closed-form
posterior distribution for
$\phi$:

$$\phi | \ldots \sim N \left( \left(\tau M^T M + T \tau_\phi I\right)^{-1} (\tau M^T R + T \tau_\phi \mu), \left(\tau M^T M + T \tau_\phi I\right)^{-1} \right)$$

After all trees are updated we can update $\tau$. The $\tau$ update comes directly from  the non-marginalised prior for $R$ combined with the prior on $\tau$:

$$\tau | \ldots \sim Ga\left( \frac{N + \nu}{2}, \frac{(y - S_1 \phi)^T (y - S_1 \phi) + \nu \lambda}{2} \right)$$

Last, we use a random-walk Metropolis-Hastings update for $\tau_\phi$, which
we found to be reliable 
for sampling from this posterior \citep{brooks2011handbook}. 
We use a standard Normal for the
proposal value to create $\tau_\phi^{*}$ and
decide to accept or reject it based on 
the acceptance probability
$\alpha(\tau_{\phi}^{*}, \tau_{\phi}) = \min \Big\{1, \frac{P(\mathbf{y}| \tau_{\phi}^{*}, \Theta) P(\tau_{\phi}^{*}) \Phi(\tau_{\phi})} {P(\mathbf{y}| \tau_{\phi},  \Theta) P(\tau_{\phi}) \Phi(\tau_{\phi}^{*})}\Big\}$, where $\Phi(x)$ represents the CDF of the standard Normal distribution. This 
is used to account for the fact that we are
using a proposal distribution with full support for a target with limited support.

\begin{algorithm}[ht]
\caption{HEBART Algorithm}\label{alg1} 

\textbf{Type:} Metropolis within GIBBS for a hierarchical BART model 

\begin{algorithmic}
\Require{\bf{y}, \bf{X}, grouping variable $\bf{z}$}
\Ensure{Posterior distribution of trees $\mathcal{T}$, $\mu$, $\phi$, $\tau$ and
$\tau_{\phi}$} 
\Init{Initial values for
$\alpha$, $\beta$, $\sigma_{\phi}$, $\tau$, number of trees P,
stumps $\mathcal{T}_{1}, \dots, \mathcal{T}_{P}$, number of observations N, number of MCMC iterations $\mathcal{I} = \text{burn-in+ post-burn-in}$, initial residual set
$\textbf{R}^{(1)} = \mathbf{y}$ }
\For{$i \leftarrow 1$ to $\mathcal{I}$}
    \For {$p \leftarrow 1$ to P}
    \begin{enumerate}
    \itemsep0em 
    \item Grow a new tree $\mathcal{T}_{p}^{*}$
  tree by either growing, pruning, changing or swapping a root node
        \item Calculate $\alpha(\mathcal{T}_{p}^{*}, 
        \mathcal{T}_{p}) = \min \Big\{1, 
        \frac{P(R_p^{(i)}| \mathcal{T}_{p}^{*}, \tau)P(\mathcal{T}_{p}^{*})}{
        P(R_p^{(i)}| \mathcal{T}_{p}, \tau)P(\mathcal{T}_{p})}
        \Big\}; $
        \item Sample $u \sim U(0, 1]$
        \item \textbf{if} $u < \alpha(\mathcal{T}_{p}^{*}, 
        \mathcal{T}_{p})$ \textbf{then do} $\mathcal{T}_{p} = \mathcal{T}_{p}^{*}$
    \end{enumerate}
    \For {$b \leftarrow 1$ to $b_p$}:      
    \text{Sample }$\mu_{b, p}$
    \For {$j \leftarrow 1$ to $J_{b_p}$}:
        \text{sample }$\phi_{b, p, j}$
    \EndFor
    \EndFor
         \State Update $\textbf{R}_{p}^{(i)} = \mathbf{y} - 
        \sum_{t\neq p}^{P} \mathbf{G(X_{ij}, z_i}, \mathcal{T}_{t}, \Theta_{t})$

     \State Update 
     $\hat f_{ij} =\sum_{p = 1}^{P}\mathbf{G(X_{ij}, z_i}, \mathcal{T}_{p}, \Theta_{p})$  
    
    \EndFor
    \State Sample $\tau$
    \State Sample $\tau_{\phi}$:
        \State \hspace{\algorithmicindent} Sample a value $d \sim N(0, \sigma^{2}_{\tau_{\phi}})$ and make $ \tau_{\phi}^{*} = \tau_{\phi} + d$
        \State \hspace{\algorithmicindent} Calculate $\alpha(\tau_{\phi}^{*}, \tau_{\phi}) = \min \Big\{1, \frac{P(\mathbf{y}| \tau_{\phi}^{*}, \Theta) P(\tau_{\phi}^{*}) \Phi(\tau_{\phi})} {P(\mathbf{y}| \tau_{\phi},  \Theta) P(\tau_{\phi}) \Phi(\tau_{\phi}^{*})}\Big\}; $
        \State \hspace{\algorithmicindent} Sample $u \sim U(0, 1]$
        \State \hspace{\algorithmicindent}\textbf{if} $u < \alpha(\tau_{\phi}^{*}, \tau_{\phi})$ \textbf{then do} $\tau_{\phi} = \tau_{\phi}^{*}$
        
 
    \EndFor
    \end{algorithmic}
\end{algorithm}

\section{Applications \& Results}\label{sec4}

This section describes the fitting of HEBART to some example data sets as we compare across different models, including:
a Linear Mixed-Effects (LME) model, fitted using the \texttt{lme4} \citep{bates2011package} package; in
\texttt{R} \citep{rcite}  and 
standard BART, fitted using the 
\texttt{dbarts} \citep{dbarts} \texttt{R} package, with the
default values for the number of trees (200), number of posterior
samples (1000), number of burn-in samples (100). 
The main performance results presented are calculated on out-of-sample values for a certain number of test sets. Since BART cannot incorporate mixed effects, other than as part of the main covariate set, we partition the results into BART including grouping information, which we expect to perform well though it does not solve the hierarchical modeling problem, versus BART without grouping information, which usually performs worse than our approach since it lacks all the data. 

The package created to run
our HEBART algorithm is available in \texttt{R} \citep{rcite}. All functions are available at \texttt{https://github.com/brunaw/hebartBase}, where the main package is stored, and \texttt{https://github.com/brunaw/hebart-experiments} contains the code
for replicating all the examples shown here,
with their corresponding
auxiliary files. 



\subsection{Simulation experiments}

We first experiment with HEBART on a simple simulation scenario, where the response
variable $y$ is simulated as a sum of a tree structure and an intra-group parameter, which
here we call $\phi_j, j = 1, \dots, 5$ and 
are simulated from a $N(0, 5^2)$ to create 
highly variant values. By splitting the simulated 
covariates $X_1, X_2, \text{ and } X_3$ at random points, we sample $n = 750$ values as

\begin{equation}
    y_{ij} \sim N\left(
    (a \hspace{1mm}\mathbf{I}(X_1 < 1)\mathbf{I}(X_2 < 0.2) +
b \hspace{1mm} \mathbf{I}(X_1 < 1)\mathbf{I}(X_2 >= 0.2) +
c \hspace{1mm} \mathbf{I}(X_1 >= 1) + d \hspace{1mm} \mathbf{I}(X_3 >= 2)) + \phi_j, 
    \tau^{-1} = 1\right),  
\end{equation}

\noindent where we have $n_j = 150$ for all 5 groups and
$(a, b, c, d)$ are randomly sampled 
from a $N(0, 3^2)$. This experiment uses
10 different train and test sets as our
cross-validation setting \citep{refaeilzadeh2009cross}. 

\begin{figure}[!h]
\begin{center}
  \includegraphics[width=120mm]{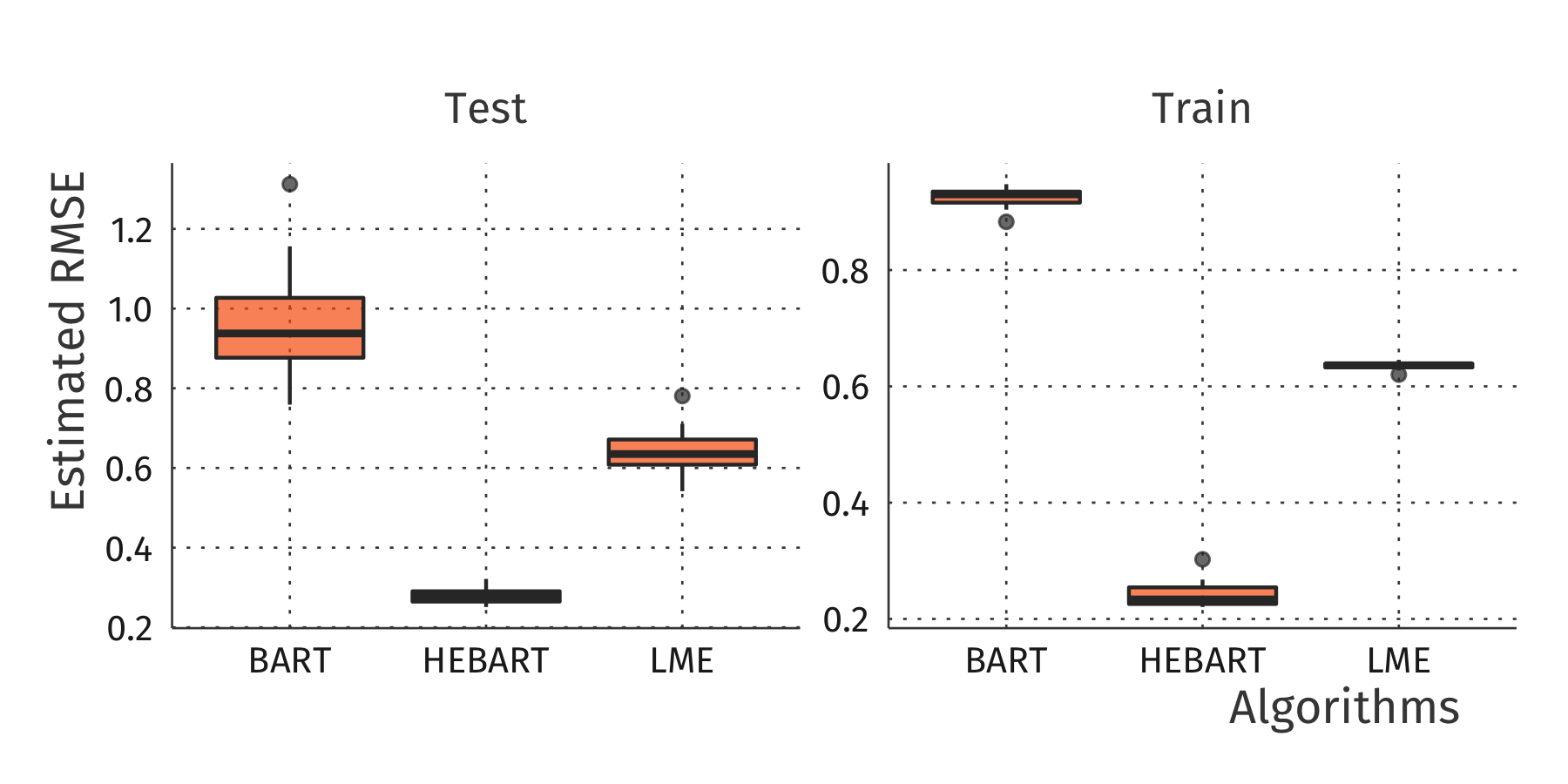} 
\end{center}
\caption{Boxplots of RMSE for testing and training sets for the 3 algorithms fitted on simulated data based on Equation (7). HEBART is the best-performing algorithm in either stage. \label{simulated_data}}
\end{figure}

To these simulated datasets, we apply three different
algorithms (LME, Standard BART and HEBART)
to the 10 training sets and make predictions for the 10 testing sets. In Figure \ref{simulated_data}, we can see that HEBART
produces the smallest root mean squared errors for both
the train and test sets, as its averages are the
closest to zero, 
followed by LME and standard BART. We also
observe that HEBART has the least varying
RMSE values on the test set, implying there
is less uncertainty around the predictions
for new data. 

In a second simulation setting 
we create $y$ with a direct HEBART
structure, by first sampling a $\mu$ value
for each terminal node, along with one
$\phi$ value for each group within the 
created terminal nodes, where 
$\tau_{\mu} = 3, \tau_{\phi} = 3$ and $\tau = 1$. 
The tree structure uses the sum of two simple 
trees, which are both based on two covariates
$(X_1, X_2) \sim \text{Normal}(0, 1)$, and uses
the previously sampled parameters to 
create the simulated response $y$. By splitting the simulated covariates $X_1, X_2$ at random points, we sample $n = 1000$ values as

\begin{equation}
    y = G(X_1, Z_1, T_1, \Theta_1) + G(X_2, Z_2, T_2, \Theta_2) + \epsilon
\end{equation}

where $\epsilon \sim N(0, 1)$. Tree $T_1$ consists of three terminal nodes with the partitions $I(X_1 < 0, X_2 < 0)$, $I(X_1 < 0, X_2 >= 0)$, $I(X_1 >= 0, X_2 >= 0)$ respectively, having terminal node values $\mu_{b_1}$ sampled from a 
$N(0, 3^{-1})$, and group terminal node values
sampled from a 
$N(\mu_{b_1}, (2/3)^{-1})$, ${b_1} = 1, \dots, 3$, as we have 3 terminal nodes. 
For tree $T_2$ the partitions are  only 
$I(X_2 < 0.5)$ and 
$I(X_2 >= 0.5)$, having terminal node values $\mu_{b_2}$
sampled from a $N(0, 3^{-1})$, and
and group terminal node values
sampled from a $N(\mu_{b_2}, (2/3)^{-1})$, ${b_2} = 1, \dots, 2$.

\begin{figure}[!h]
\begin{center}
\begin{tabular}{cc}
  \includegraphics[width=120mm]{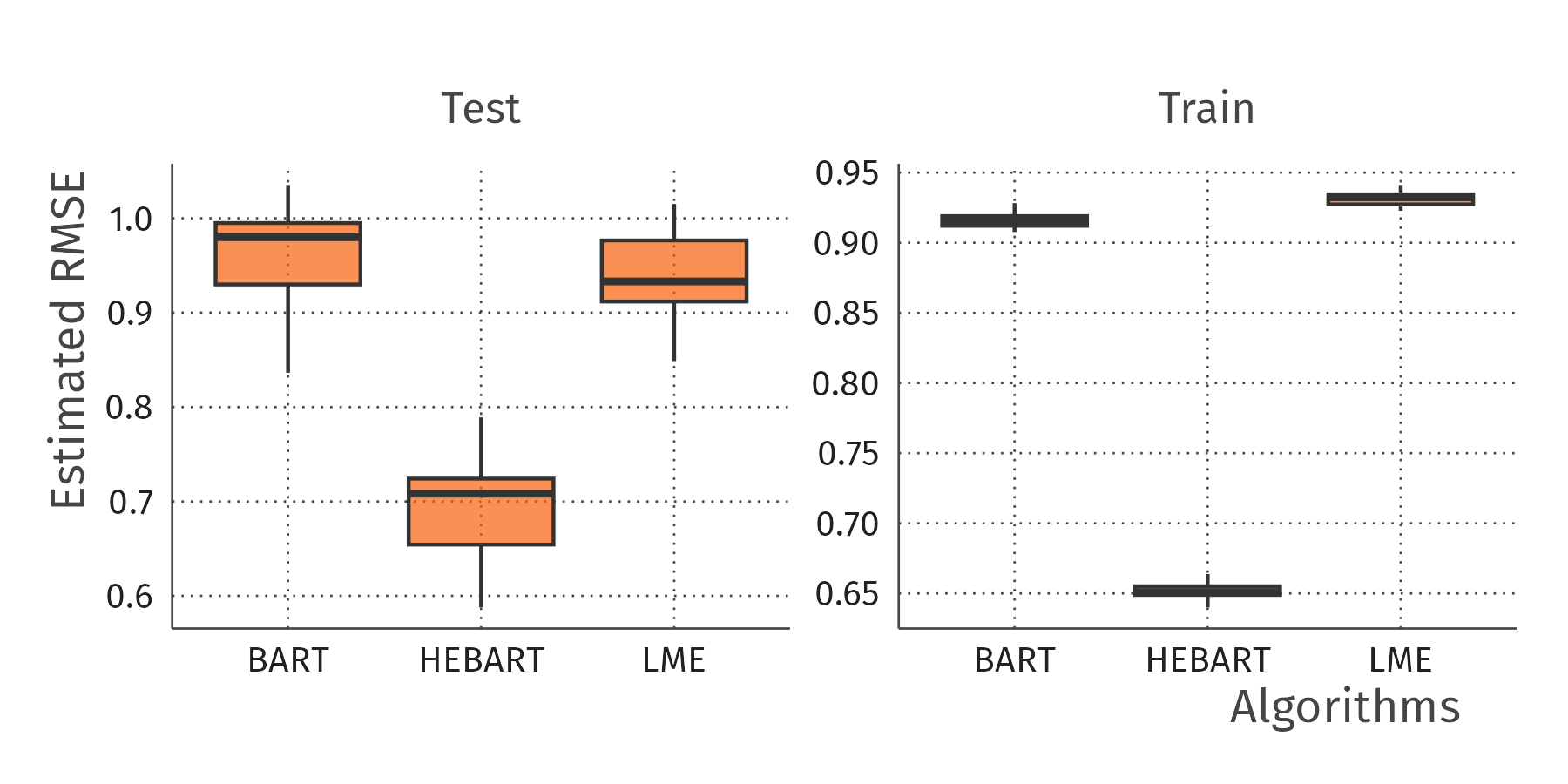} \\
 (a) Boxplots of RMSE for testing and training sets for the 3 algorithms fitted on simulated data \\[6pt]
 
  \includegraphics[width=100mm]{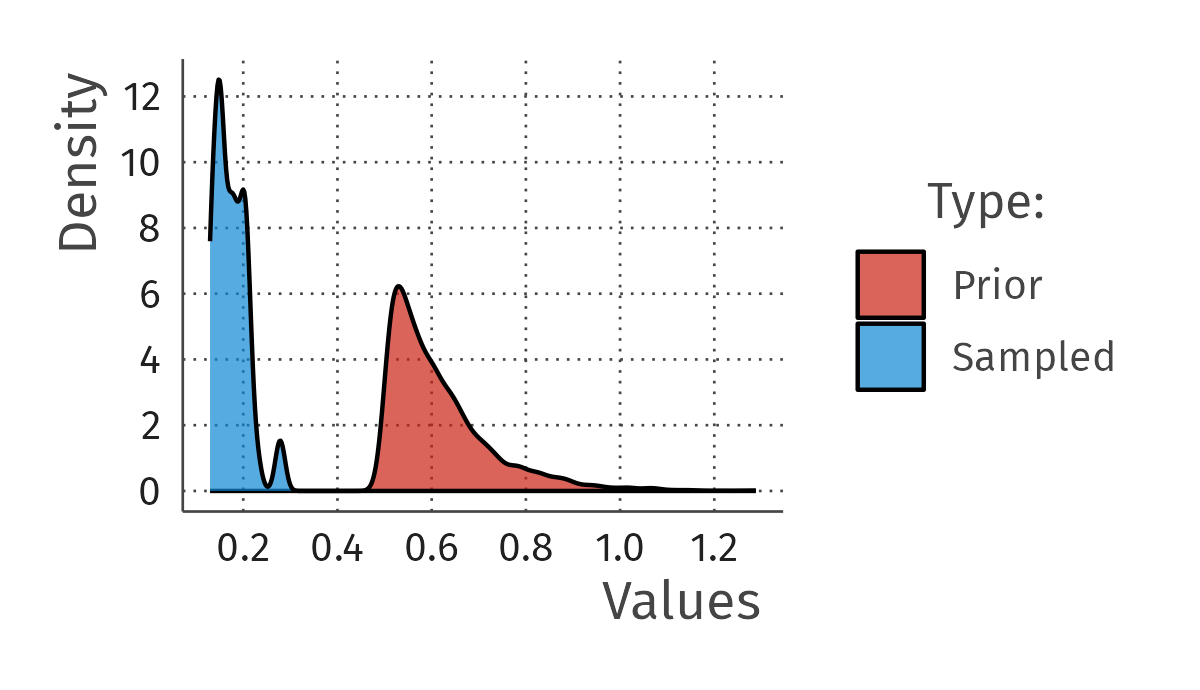} \\
 (b) Prior and posterior densities for $\tau_{\phi}$ for one run of HEBART \\[6pt]
 
\end{tabular}
\end{center}
\caption{
(a) For RMSE, HEBART exhibits superior performance in the test and train sets compared to its two competitors for the second simulation setting; (b) The prior
and posterior densities show how the
two distributions significantly
differ in this case. It appears that the information in the
data pushes the posterior of $\tau_\phi$ to be lower than indicated by the prior distribution obtained from a random intercepts model fit. 
\label{simulated_second}}
\end{figure}

The results of this simulation are shown in
Figure \ref{simulated_second}. Once again,
we observe HEBART to be the best-performing algorithm, as its test RMSE average is lower than its competitors. We also include here the prior
and posterior densities of $\tau_\phi$ one of the runs of the HEBART model. The two densities
are markedly different, with the posterior having a distinctly larger variance (lower precision) associated with the grouping information. This is not necessarily surprising given that the grouping information for this simulation scenario is specified at the sub-terminal node level rather than through an additive effect as is used in an LME.

\subsection{Real data sets}

\subsubsection{Sleep study data}\label{sleepdata}

This dataset 
consists of the first 10 days of a sleep study
\citep{belenky2003patterns}. The response variable is the 
average reaction time per day (in milliseconds), and the
covariate is the number of days of sleep deprivation, 
for 18 subjects. Linear mixed-effect models 
\citep{pinheiro2000linear} are
often applied to this data \citep{bates2011package}, 
so we can directly compare our results. 

We create a 20-fold cross-validation and evaluate the model on the left-out data.
Our HEBART model is set to use 10 trees, has hyperparameters $\alpha$ = 0.95 and $\beta$ = 2, and
all the other prior hyperparameters are set as 
described above. For the MCMC sampling, we use 1500
iterations with 250 iterations of burn-in. We also
fit an LME \citep{pinheiro2000linear}
and a standard
BART model \citep{chipman2010bart} to compare our
results. Our main interest is in the performance of the predictions and the estimate of the group-level random effect standard error, though this is not available in the BART model since it has no such parameter. However, as described above the grouping variable is sub-optimally included as one of the covariates. 

\begin{figure}[!h]
\begin{center}
\begin{tabular}{cc}
  \includegraphics[width=153mm]{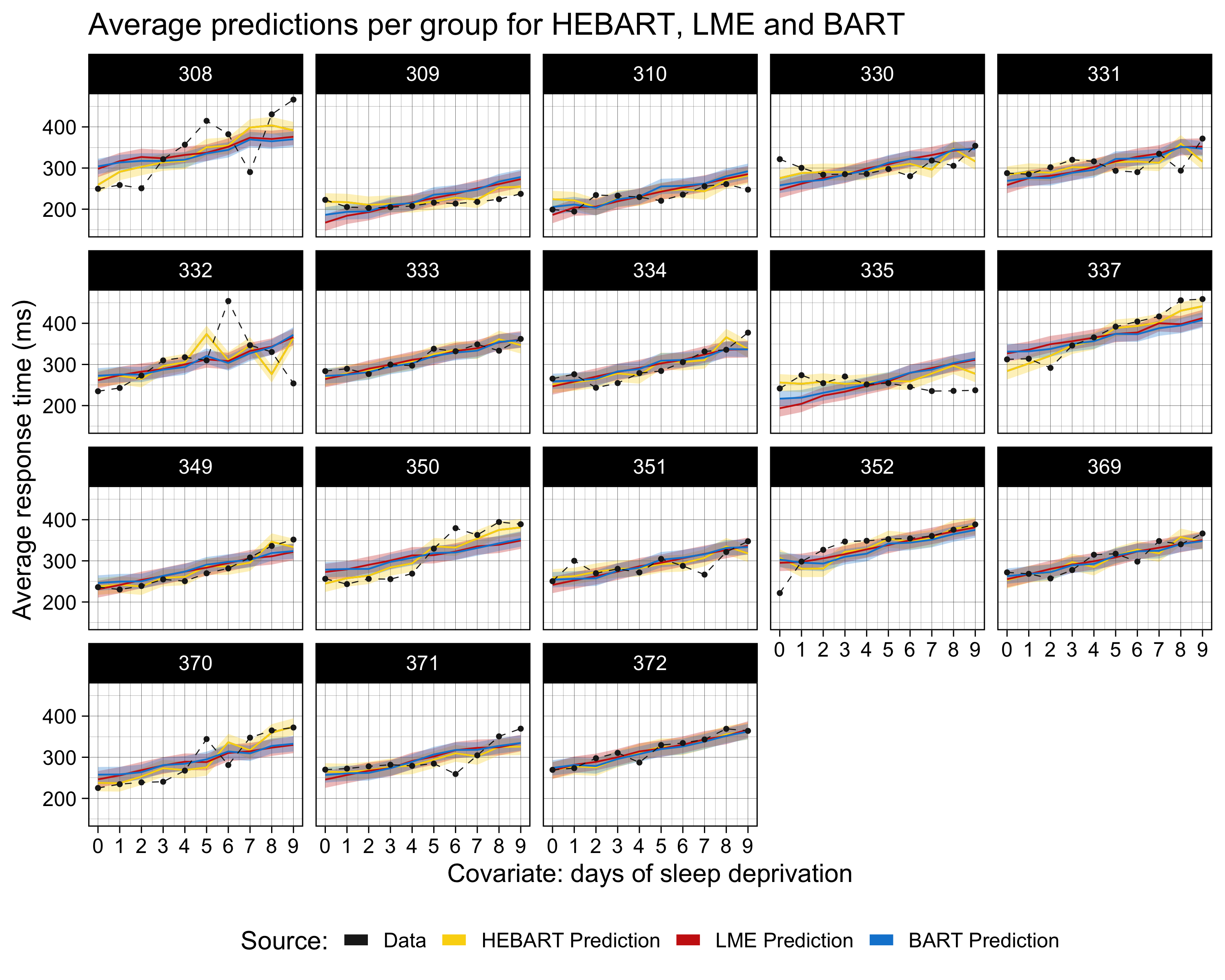} \\
 (a) Average predictions for the groups in the sleep study dataset \\[6pt]
\end{tabular}
\end{center}

\begin{center}
\begin{tabular}{c|c|c|c|c}
Source & HEBART & LME & BART+Group \\ 
\hline 
Test set &  27.7 [8.22, 47.2] & 32.4 [15.8,49]  & 32.3 [13.3, 51.2] \\
Train set & 16.9 [10.1,23.7] & 29.3 [28.3,30.3] & 28.1 [26.6,29.6] 
\\
\vspace{3mm}
\end{tabular}
\\ 
\begin{tabular}{c}
(b) Average train and test set RMSE values, with the corresponding empirical 95\% confidence intervals \\[6pt]
\end{tabular}
\end{center}

\caption{(a) Average predictions for each patient ID 
in the sleep study data, using HEBART, LME and BART
with confidence/credible intervals. HEBART produces 
predictions closer to the actual true data, including the most challenging IDSs; 
(b) Average test RMSE values, with empirical 95\% confidence intervals. The lowest values are  for HEBART. 
\label{sleepstudy_data}
}

\end{figure}

In Figure \ref{sleepstudy_data} we show the 
HEBART, BART and LME
predictions for each group ID, for predictions created when those
observations were not included in the training set. In
other words, we are seeing the average predictions for 
the test sets, split by group ID. The plot also
shows the true observations as green dots for comparison. We can see for almost all IDs, 
the HEBART predictions are the ones closest to the
true values, especially for more difficult cases,
such as IDs 335, 332, 351 and 351. Overall, 
HEBART is able to adapt better to changes in each 
of the individual group patterns. In the same 
Figure, we also have the RMSE table for each model, 
with their corresponding empirical 95\% confidence
intervals, calculated using the results from the 20
train and test sets. From this table
it is clear that HEBART produces the 
best results, as both the train and test
RMSEs are the lowest of the three methods, with similarly low values for the confidence intervals. 

\subsubsection{Gapminder data}\label{sec_gapminder_data}

Another standard dataset used to
exemplify mixed effects modeling is the Gapminder data \citep{lang2011gapminder}. 
For our experiment, the subset of
the data consists of the life expectancy 
values (in years) for 20
different countries, from 1950 to 2018. We use life expectancy as the response and country as the grouping variable with year as the sole covariate. We separate the dataset
into 10 different training and test sets, 
where the testing set is composed 
of all the observations for 15 sampled
years (for all countries), and the 
corresponding training set is composed
of the observations for the remaining data.
The idea behind this is, for 
each resample, to fully remove a set 
of years from the training set to 
make it harder for the model to predict 
for such years, since it has no information about what happened in 
the removed years. We also compare our 
model against LME, BART
and BART using the country as a covariate. We expect the latter to perform well since it has all the information, but we remind the reader that this method does not have the added advantage of HEBART in being able to predict with fully or partially missing grouping values. 

\begin{figure}[!h]
\begin{center}
\begin{tabular}{cc}
  \includegraphics[width=120mm]{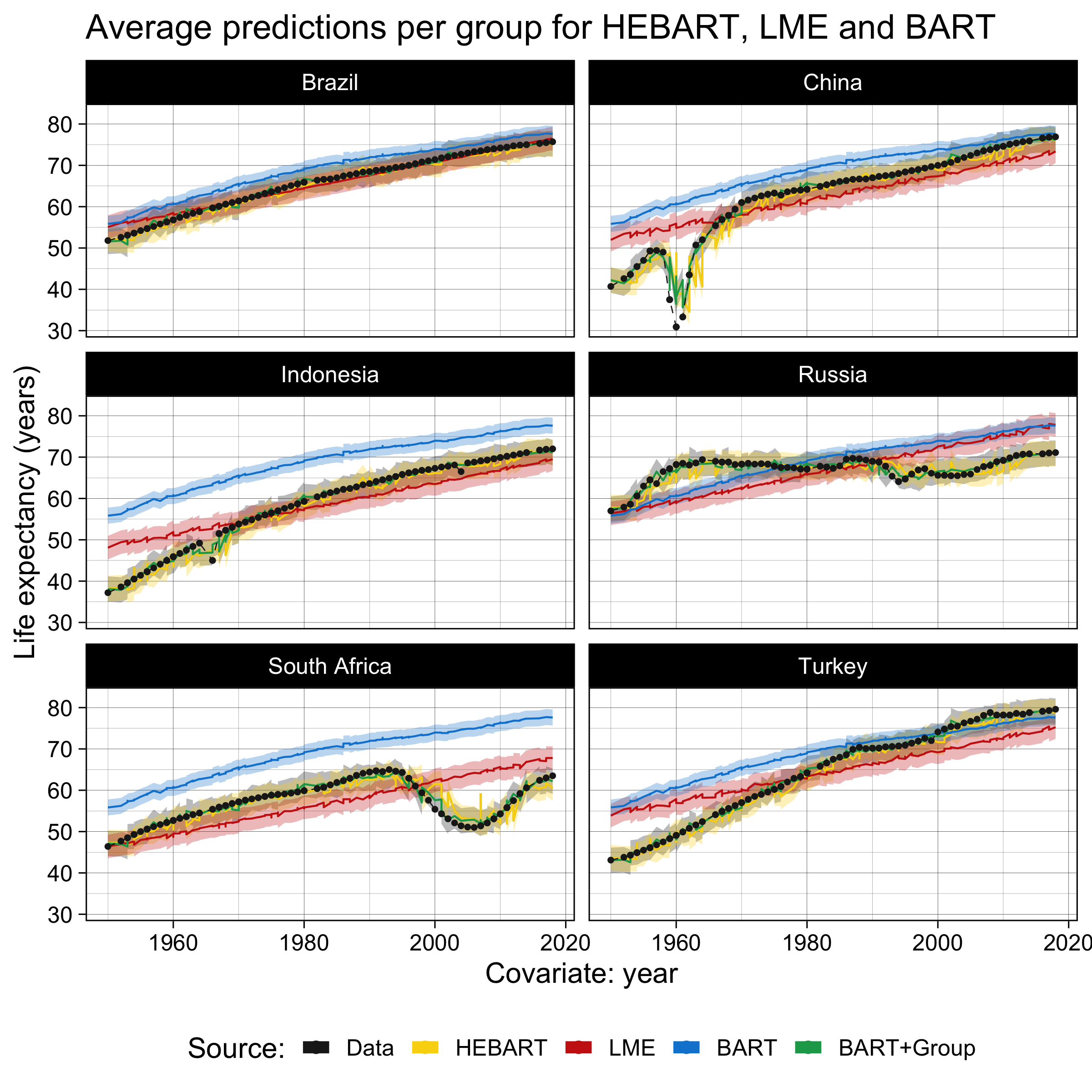} \\
 (a) Average out of sample predictions for selected countries in the gapminder dataset \\[6pt]
\end{tabular}
\end{center}

\begin{center}
\begin{tabular}{c|c|c|c|c}
Source & HEBART & LME & BART & BART+Group \\ 
\hline 
Test set &  1.33 [0.77, 1.88] & 3.59  [2.84, 4.33]  & 7.53 [6.53, 8.52] & 0.815 [0.43, 1.20] \\
Train set & 0.793 [0.162, 1.42] & 3.57  [3.38 , 3.77] & 7.47  [7.20, 7.73] & 0.351 [0.310, 0.391]
\\
\vspace{3mm}
\end{tabular}
\\ 
\begin{tabular}{c}
(b) Average training and test RMSE values, with the corresponding empirical 95\% confidence intervals \\[6pt]
\end{tabular}
\end{center}

\caption{(a) Average predictions (considering the 10 test sets) for 6 selected countries in the gapminder dataset, 
using HEBART, LME, BART, and BART using group as a covariate, 
with confidence/credible intervals. HEBART and BART+Group both
produce predictions closer to the actual true data for all cases
shown; 
(b) Average test RMSE values, with empirical 95\% confidence intervals for the mean RMSE. The table confirms the lowest values for HEBART, with neither the training value not even
intersecting with the other ones, except for HEBART and LME. 
\label{gapminder_data}}
\end{figure}

Figure \ref{gapminder_data} shows the
results for this experiment as an 
RMSE table and the depiction of the 
predicted values in comparison to the 
true observations. Note that we have 10 training
and test sets, and what we show in the Figure 
are the average predictions for the 10 test sets. 
So, when we bind all the predictions 
for all the 10 test sets, we have at least
one prediction for each year of all countries. 
As above,
the HEBART prediction
is very close to the actual values due to the increase in  flexibility
of the predictions from the non-parametric structure. The performance of the BART model that uses
the country as a covariate (BART+Group) is slightly superior but as stated above, this is not unexpected since the 
BART+Group misses the fundamental advantage of HEBART in separating out (and potentially removing) the country-level variability. Our model, on the 
other hand has the advantage of being able
to predict when grouping variable information
is missing, but still retains the prediction performance. 

\section{Conclusions}\label{sec5}

We have provided a new extension of Bayesian Additive Regression Trees that allows for structured data to be used appropriately at the model fitting and prediction stage. Where a grouping variable is present, we are able to provide predictions at both the group level and the level above it. The flexible tree structure thus allows us to produce excellent predictions without the need to specify how the grouping variable is included in the model structure. However, we still retain the ability to report and remove the group-level variability by having a parameter that represents the group-level standard deviation. 

In simulation-based and real data studies we have shown that the model performs better than other common linear and mixed effects approaches or BART itself. Our approach is not comparable to many other standard Bayesian machine learning methods due to their inability to handle the grouping information other than as a na\"ive covariate. In the real data examples of Section \ref{sleepdata} and \ref{sec_gapminder_data} we have demonstrated that HEBART performs well against LME modelling strategies despite these being the archetypal examples of datasets where linear mixed-effects models would fit well. Finally, We see many potential extensions of our approach, 
including: (1) extending the hierarchical data structure to multiple or nested grouping variables; (2) explicitly modeling joint random effects using the multivariate normal distribution and so estimating covariances between grouping variables; (3) including recent
BART extensions, such as SOFT-BART
\citep{linero2018abayesian} and MOTR-BART
\citep{prado2021bayesian} which can substantially enhance the predictive capabilities of the model; 


\section*{Acknowledgments}
 
\footnotetext{\textbf{Abbreviations:} BART, Bayesian Additive Regression Trees; HEBART, Hierarchical Embedded Bayesian Additive Regression Trees; MH, Metropolis-Hastings; MCMC, Monte Carlo Markov Chain; RMSE, Root Mean Squared Error; LME, Linear Mixed-Effects;}

\section*{Acknowledgments}
 
This work was supported by the Science Foundation Ireland Career Development Award grant number: 17/CDA/4695. Andrew Parnell’s work was also supported by: a Science Foundation Ireland investigator award (16/IA/4520); a Marine Research Programme funded by the Irish Government, co-financed by the European Regional Development Fund (Grant-Aid Agreement No. PBA/CC/18/01); European Union’s Horizon 2020 research and innovation programme InnoVar under grant agreement No 818144; SFI Centre for Research Training in Foundations of Data Science 18/CRT/6049, and SFI Research Centre awards I-Form 16/RC/3872 and Insight 12/RC/2289\_P2. For the purpose of Open Access, the author has applied a CC BY public copyright licence to any Author Accepted Manuscript version arising from this submission. 

\subsection*{Author contributions}

Bruna Wundervald has contributed to 
the project ideation, mathematical development, 
code implementation, production and
evaluation of results. 
Andrew Parnell has contributed to 
the project ideation, mathematical development, 
code implementation, and
evaluation of results. 
Katarina Domijan has contributed to 
the project ideation, mathematical development, and
evaluation of results. 

\subsection*{Financial disclosure}

This work was supported by the Science Foundation Ireland Career Development Award grant number: 17/CDA/4695.

\subsection*{Conflict of interest}

The authors declare no potential conflict of interest.

\bibliography{main.bib}

\nocite{*}

\end{document}